\documentclass[twocolumn,spreprintnumbers,amsmath,amssymb]{revtex4}
\usepackage{graphicx}

\newcommand{\ee}{\end{eqnarray}}
\newcommand{\be}[1]{\begin{eqnarray} \mbox{$\label{#1}$} }
\newcommand{\mtrx}[2]{\left(\begin{array}{#1} #2 \end{array}\right)}
\newcommand{\eref}[1]{(\ref{#1})}
\newcommand{\der}{\mathrm{d}}

\newcommand\eg {{\it e.g.~}}

\newcommand\grad{\vec\nabla}

\newcommand{\ul}\underline

\newcommand{\pr}{^\prime}

\renewcommand{\vec}[1]{\text{\boldmath{$ #1 $}}}
\newcommand{\uv}[1]{\vec{\hat{#1}}}

\newcommand{\fermi}{\mathrm{F}}

\newcommand{\erg}{\epsilon}

\newcommand{\ordo}[1]{\mathcal{O}(#1)}
\newcommand{\elec}{c}
 
\newcommand{\vect}[1]{\vec{\tilde{#1}}}
\newcommand{\up}{\uparrow}
\newcommand{\down}{\downarrow}
\newcommand{\Gammaf}{\Gamma_\mathrm{f}}
\newcommand{\Gammas}{\Gamma_\mathrm{s}}
\newcommand{\chcon}{\mathcal{C}}
\newcommand{\parity}{\mathcal{P}}
\newcommand{\timerev}{\mathcal{T}}
\newcommand{\omegac}{\omega_\mathrm{c}}
\newcommand{\texpf}{\zeta}
\newcommand{\texps}{\eta}
\newcommand{\kbol}{k_\mathrm{B}}

\newcommand{\jelt}{\alpha}
\newcommand{\jhe}{\gamma}
\newcommand{\jht}{\kappa}

\begin{document} 

\title{A Nernst effect beyond the Sondheimer formula} 

\author{Janik Kailasvuori
%$^1$ and  Stefan Kirchner$^{1,2}$
}
%\email {kailas@pks.mpg.de}

\affiliation{
%$^1$
Max-Planck-Institut f\"ur Physik komplexer Systeme, N\"othnitzer Stra\ss{}e 38, 01187 Dresden, Germany
% \\
%$^2$Max-Planck-Institut f\"ur Chemische Physik fester Stoffe, N\"othnitzer Stra\ss{}e 40, 01187 Dresden, Germany
}

\date{\today}

\begin{abstract} 
\noindent
The Nernst effect is known to be a sensitive probe of the scattering mechanisms.  Large  Nernst effects, a decade ago a topic mainly in vortex dynamics in superconductors, have recently attracted interest in correlated systems like  cuprate metals  and  heavy fermion systems. However, it has been found that these ``giant" Nernst signals stay within one order of magnitude of  the classical Sondheimer formula for a Fermi liquid.  It therefore seems that a large Nernst effect neither  relies on nor is sensitive  to strong correlations. This might, however,  be different in some recently studied supposedly correlated systems like FeSB$_2$. We do not address these systems specifically, but  take them  only as one inspiration for considering a theoretical scenario where  correlations could actually  sizably modify the Sondheimer formula. We point out that  an enhanced Nernst signal would follow from the phenomenological Boltzmann approach introduced by Coleman, Schofield and Tsvelik\cite{Coleman-Schofield-Tsvelik-PRL-1996} proposed to capture electronic correlations by imposing different relaxation times on the charge conjugation even and odd Majorana components of the electron. 

 \end{abstract} 

\maketitle

\noindent
Thermoelectric effects are interesting for the information they provide about a conductive system, but also for practical and commercial applications. One such application would be Peltier cooling, where a heat current can be directed with the help of electric fields. Inversely, waste heat could be transferred into electricity. The thermoelectric efficiency in materials to date meets  commercial requirements only at high temperatures of several hundred Kelvin, whereas finding commercial compounds at cryogenic temperatures remains a challenge.
% although impressively high in some compounds like the Bi$_2$Te$_3$-Sb$_2$Te$_3$ alloy. 
Further understanding and modeling is therefore desirable. 

A thermoelectric coefficient of big importance for the thermoelectric efficiency is the thermopower, a.k.a. Seebeck coefficient. It appears in the figure of merit $Z T=\tfrac{S^2 \sigma_0}{\kappa} T$, which should exceed 1 for commercial applications of Peltier cooling. 
The interest in electronic correlations as a source of exceptional thermoelectric properties has been intensified by the observation of large thermopowers in semiconductors of transition metal compounds such as FeSi \cite{Sakai-Ishii-Onose-Tomioka-Yousuhash--Adachi-2007}, Na$_x$CoO$_2$ \cite{Terasaki-Sasago-Uchinokura-1997} and FeSb$_2$ \cite{Bentien-Johnsen-Madsen-Iversen-Steglich-2007}.  FeSb$_2$ displays an impressive thermopower of $S=-45 \, \textrm{mV/K}$ at 10~K\cite{Bentien-Johnsen-Madsen-Iversen-Steglich-2007}. 
 It has been conjectured that such enhanced thermopowers would be due to electronic correlations \cite{Bentien-Johnsen-Madsen-Iversen-Steglich-2007, Sun-Oeschler-Johnsen-Iversen-Steglich-PRB-2009,Sun-Oeschler-Johnsen-Iversen-Steglich-APE-2010,Sun-Oeschler-Johnsen-Iversen-Steglich-DALTON-2010}. Some model calculations \cite{Schweitzer-Czycholl-1991,Palsson-Kotliar-1998,Oudovenko-Palsson-Haule-Kotliar-Savrasov-2006,Grenzebach-Anders-Czycholl-Pruschke-2006} give indeed support for this. 
A very recent model calculation \cite{Tomczak-etal-Kotliar-2010}  accounting for correlations effect finds some qualitative agreement with the experiments on FeSb$_2$, but also some strong quantitative disagreement, and it is speculated that also phononic contributions might play an important role.

Another thermoelectric coefficient of particular informativeness and closely related to the thermopower is the Nernst signal \cite{Ettinghausen-Nernst-1886,Seeger-book-1990}, the transverse electric field as a response to a  temperature gradient in a system  with an out-of-plane magnetic field and with zero charge current.  In ordinary electron systems the Nernst signal is  known to be  a sensitive probe of the scattering mechanism.   Just like the thermopower related Peltier effect  can be used for thermoelectric cooling, so can also the Nernst effect related Ettinghausen cooling be used.
%However, the typical figures of merit for the latter are typically smaller. [REFS. DISCUSS MORE WITH PEIJIE] 

In the 1990s the Nernst effect was  a subject mainly in high-$T_\mathrm{c}$ in the superconducting regime,  where it is the transverse motion of vortices in a temperature gradient that effectively leads to a transverse charge flow and thus a transverse voltage.   
Surprisingly high Nernst signals not due to vortex transport has recently been observed in cuprates  above $T_\mathrm{c}$  in the pseudogap state \cite{Xu-Ong=Wang-Kakeshita-Uchida-Nature-2000}  (see ref.~\cite{Behnia-review-2009} for further references) as well as in other semiconductors and metals with non-standard Fermi liquid behavior.\cite{Bel-Jin-Behnia-Flouquet-Lejay-PRB-2004,Behnia-Measson-Kopelevich-PRL-2007} This lead to   the idea that a "giant" Nernst effect would be related to strong correlations. 
 This also sparked renewed interest in the effect in Fermi liquid systems, which before this century attracted little attention, among other because in metals, with the approximate electron-hole symmetry around a high Fermi surface, the signal is small, being zero in a perfectly electron-hole symmetric case due to Sondheimer cancellation. One exception is the poor metal bismuth (Bi). There the large Nernst effect, measured already in the late 19th century \cite{Ettinghausen-Nernst-1886}, surpasses the one measured more recently in more exotic electron systems like heavy fermion systems, overdoped cuprates and organic superconductors.\cite{Behnia-Measson-Kopelevich-PRL-2007} 

The Nernst effect in bismuth, although astonishingly large, complies at low temperatures (up to 5 K) nonetheless within one order of magnitude \cite{Behnia-Measson-Kopelevich-PRL-2007} with the formula for the Nernst effect of  a degenerate Fermi system derived by Sondheimer in 1948 \cite{Sondheimer-1948}. This seems to kill the idea that strong correlations would be necessary for a giant Nernst effect.  Furthermore,  Behnia and coworkers \cite{Behnia-Measson-Kopelevich-PRL-2007,Behnia-review-2009}  point out that also the correlated systems with a Nernst effect characterized as ``giant" are  in their respective low-temperature regimes $T\ll T_\fermi$  fitted, within one order of magnitude, by the Sondheimer formula.  The situation for the thermopower is the same in that respect that its ratio to the specific heat is of the same magnitude as the one for an electron gas.\cite{Behnia-Jaccard-Floquet-2004}
Thus, it seems that  the presence of strong correlations is neither a necessary nor even a sufficient condition for observing Nernst effects that would be truly anomalous and hence that the Nernst effect is essentially insensitive to correlations.

This situation seems to be different in recently studied semiconductors with large Nernst signals like FeSb$_2$,  with a Nernst signal almost as big as the low-temperature value for bismuth. As already mentioned, this material shows indications of strong  correlations. Furthermore, Hall response data appear to  support a single-band description of the electronic properties.\cite{Sun-private} An analog of the Sondheimer formula for a single-band degenerate quantum gas  can also be derived in the opposite nondegenerate limit, applicable to semiconductors at low enough temperatures.  The resulting formula has the same sensitivity to the energy dependence of the scattering mechanism.  
Applying it to the Nernst effect of FeSb$_2$ as a function of temperature, however, does not give the pertinent Sondheimer result even up to an order of magnitude and appears to indicate scattering mechanisms of highly awkward energy dependences \cite{Sun-private}. One  scenario is that the formula remains valid, but some special feature has not been accounted for. This could be anomalous features in the density of states, for example due to resonant levels.\cite{Heremans-Jovovic-etal-Science-2008} This could also be the lacking account of phonon drag contributions, although  
for FeSb$_2$ phonon drag effects have been argued to be small below 10 K.\cite{Sun-Oeschler-Johnsen-Iversen-Steglich-PRB-2009}
Still another scenario is that the standard formula actually breaks down due to electronic correlations. 

In this paper we are interested in the  scenario that electronic correlations could modify substantially the Sondheimer formula. However, the specific example of Nernst effect in undoped FeSb$_2$ is only a general inspiration to us but will in no way be addressed by our study. For once, our model is too general and simplistic, but most of all,  we need to assume a degenerate Fermi system rather than the more relevant non-degenerate one. As a very speculative remark we find it interesting to mention that the tellurium doped derivative  Fe(Sb$_{1-x}$Te$_x$)$_2$ with $x=0.01$ displays instead metallic behavior \cite{Sun-private}.  At the moment of writing we do not know of Nernst effect measurements in this doped compound. However, for this case our study could in principle be interesting, would there be any deviations from the Sondheimer formula due to correlations, at least as long as we are dealing with the minority class of systems amenable to a one-band description.

The idea we want to explore is based on the work by Coleman, Schofield and Tsvelik (CST) \cite{Coleman-Schofield-Tsvelik-PRL-1996, Coleman-Schofield-Tsvelik-1996}. Inspired by a conjecture by Anderson \cite{ Anderson-1991-tworelaxationtimes} they proposed a model for explaining the apparent but perplexing occurrence of two different relaxation times needed to interpret the temperature dependences of the observed longitudinal and Hall components of the conductivity in cuprate superconductors in their normal metallic state. Criticizing existing proposals, like the one based on 'hot-spot scenarios' or skew-skattering, for working only in fortuitous circumstances, they opted for a model  based on more general principles. Pointing out  that the longitudinal and the  Hall current transform differently only with respect to  charge conjugation ($\chcon$) but neither under parity ($\parity$) nor under time reversal ($\timerev$), they suggested looking at a theory written in terms of $\chcon$-even and $\chcon$-odd Majorana superpositions of electrons and holes,
 both with diagrammatics as well as with a phenomenological Boltzmann analysis, by ad hoc imposing different relaxation times for the $\chcon$-even and $\chcon$-odd Majorana components. These phenomenological relaxation times were for simplicity assumed to be constant in energy but allowed to be of different power in their temperature dependences. From the observed temperature dependences and assuming one relaxation time to be much shorter than the other one, CST could derive the electrothermal transport tensors and deduce their respective temperature dependencies with an impressing although not complete matching with experiments. The microscopic mechanism underlying this separation of relaxation times remain an open question but, like Anderson, CST speculate in that it could be related to strong correlation effects like spin-charge separation. We will return to the model of CST further down, but before that we now introduce the thermoelectric transport coefficients.

In the diffusive regime the  charge current $\vec{j}^\textrm{el}$ and the heat current $\vec{j}^\textrm{h}$ driven by an electrochemical field $\vec{\mathcal{E}}=\vec{E}-\tfrac{1}{e}\nabla \mu$ and a temperature gradient $\nabla T$ are given by \cite{Seeger-book-1990}
\be{transcoff}
\vec{j}^\textrm{el} & = & \bar{\sigma} \vec{\mathcal{E}} - \bar{\jelt} \grad T 
\\ 
\vec{j}^\textrm{h} & = &  \bar{\jhe} \vec{\mathcal{E}} - \bar{\jht} \grad T \, .
\ee
The conductivity tensor $\bar{\sigma}$ has in an isotropic system the longitudinal components $\sigma_{xx}=\sigma_{yy}\equiv \sigma_0$ and the transverse components $\sigma_{xy}= -\sigma_{yx}$, the latter linear in the magnetic field $\vec{B}=B \uv{z}$. The same structure applies for the other transport tensors $\bar{\jht}$, $\bar{\jelt}$ and  $\bar{\jhe}$.  The Onsager relation  $\bar{\jhe}(\vec{B})=T\bar{\jelt}^\mathrm{t}(-\vec{B})$  reads in our case simply $\bar{\jhe}=T\bar{\jelt}$. For degenerate electron gases in a single band  one derives the Wiedemann-Franz relation $
{\kappa}/{\sigma_0} = T \mathcal{L}_0 $, in terms of the Lorentz number $\mathcal{L}_0={\pi^2 \kbol^2}/{3 e^2}$. One can also define the Hall Lorentz number $\mathcal{L}_{xy}\equiv \kappa_{xy}/ T\sigma_{xy}$, which equals $\mathcal{L}_0$ for an electron gas.

The Seebeck coefficient $S$ (a.k.a. the thermopower $Q$) and the Nernst signal $N$ address the electric field $\vec{E}=\bar{\sigma}^{-1} \bar{\jelt}\nabla T=\bar{Q} \nabla T$ induced by a  thermal gradient in the absence of an electric current $\vec{j}^\textrm{el}$. Here $\bar{Q}$ is the thermopower tensor. With the gradient $\nabla T $ purely in the $x$-direction (isothermal case) they are 
\begin{eqnarray}
S/e\equiv Q_{xx} \equiv \frac{\sigma_{xx}\jelt_{xx}+\sigma_{xy}\jelt_{xy}}{\sigma_{xx}^2 +\sigma_{xy}^2} \label{seebeck}
\\ 
N \equiv Q_{yx}=\frac{ \sigma_{xy} \jelt_{xx} - \sigma_{xx} \jelt_{xy} }{\sigma_{xx}^2 +\sigma_{yy}^2}\, . \label{nernst}   
%= e\mathcal{L}_0 \left (\frac{\der \log \sigma(\erg)}{\der \erg}\right )_{\erg=\erg_\fermi}   \, .
\end{eqnarray}
(Concerning the sign convention $N=+E_y/\partial_x T$, see \cite{Behnia-review-2009}.)

For the case of a degenerate electron gas it is convenient to introduce $\sigma(\erg)=e^2 \tau(\erg) \int_\vec{k} \delta (\erg-\erg_\vec{k}) v_\vec{k}^2/d$, which satisfies $\sigma (\erg_\fermi)=\sigma_0$. Here, $d=2,3$ is the dimension. With the proper generalization for the Hall component $\sigma_{xy}$ and using that for the degenerate case 
$\bar{\jelt}= \tfrac{\der}{\der \erg}(\bar{\sigma})$, introducing $(\ldots)\pr \equiv \tfrac{\der}{\der \erg}(\ldots)|_{\erg=\erg_\fermi}$, 
one finds in the limit of a small magnetic field ($|\sigma_{xy}|\ll \sigma_{xx}$ and $|\jelt_{xy}|\ll \jelt_{xx}$) the Mott formula 
\be{mott}
S/e \mathcal{L}_0 =\left ( \log \sigma_{xx}\right )\pr = \left ( \log \mu \right )\pr + \left ( \log n \right )\pr  
\ee
and the Sondheimer formula  (more precisely a low temperature limit thereof \cite{Sondheimer-1948}\footnote{The formula presented in ref.~\cite{Sondheimer-1948} is not accurate. Accurate formulations can be found in textbooks like ref.~\cite{Seeger-book-1990}.})
\be{nernst1} 
N/e\mathcal{L}_0= -   \left(\frac{\sigma_{xy}}{\sigma_{xx}}\right)\pr = -B (\mu_\mathrm{H})\pr  \, .
\ee 
where $\mu_H=r_\mathrm{H} \mu $ is the Hall mobility, related to the mobility through the Hall factor $r_\mathrm{H}$. The rewritings to the right in \eref{mott} and \eref{nernst1} stress that whereas the Seebeck coefficient draws its  energy dependence both from the mobility and the density, the Nernst signal depends only on the energy dependence of the (Hall) mobility. 
%^(Contributions from $(m^*)\pr$ have been neglected in both cases.)
  In contrast to the sign of the Hall conductivity $\sigma_{xy}$, which depends on the sign of the charge carriers, 
the sign of the Nernst signal is independent of charge carrier sign, but is on the  other hand sensitive to the energy dependence of the scattering mechanism. Assuming $\tau(\erg)=\tau_0\erg^r$, where $r\sim \ordo{1}$ for ordinary scattering mechanisms \cite{Seeger-book-1990}, one can for the single band case derive the Nernst coefficient $\nu\equiv N/B$ 
\begin{eqnarray} 
\nu\, = &-\frac{|e|\mathcal{L}_0  T \mu_\textrm{H}}{ \epsilon_F } r &\quad \textrm{degenerate} \label{nernstdeg} \\
\nu \,= & -\frac{k_\textrm{B} \mu_\textrm{H}}{|e|}  r &\quad \textrm{nondegenerate} \label{nernstnondeg} \, .
\end{eqnarray}
The quantity $\nu/\mu_\mathrm{H}$ can consequently be used to extract the energy dependence of the relaxation time. 
The exponent $r$ is for standard types of scattering of the order of 1 (\eg  $r=+3/2$ for ionized impurity scattering and $r=-1/2$ for acoustic deformation potential scattering). In the particle-hole symmetric case $r=0$ the Nernst effect vanishes, which is known as Sondheimer cancellation.  This should be contrasted with $S =e\mathcal{L}_0 (3/2+r) T/\erg_\fermi $ for the degnerate ( $S=(k_\mathrm{B}/e)(\tfrac{5}{2}+r-\erg_\fermi/k_\mathrm{B}T )  $   for the nondegenerate) case \cite{Seeger-book-1990} where the scattering potential only plays a minor role as a consequence of the density contribution in the Mott formula \eref{mott}.  The formulae \eref{nernstdeg} and \eref{nernstnondeg} apply only for a single band. For the case when both electrons and holes contribute  one can derive a more complicated formula, see \eg ref.~\cite{Seeger-book-1990}, 
%\be{}
%\nu &=& - \frac{k_\textrm{B}}{e} \left[  
 %\left(
%\frac{\sigma_n}{\sigma} \mu_{\textrm{H}n}+\frac{\sigma_p}{\sigma} \mu_{\textrm{H}p} \right )\, r 
%+\right .   \nonumber \\
%& & \left. +
 %\frac{\sigma_n\sigma_p}{\sigma}
 %(  \mu_{\textrm{H}n}-\mu_{\textrm{H}p} ) (2 r +5 +\erg_\textrm{gap}/k_\textrm{B}T)
%\right ]
%\ee
%when both electrons (n) and holes (p) contribute, however, 
in which the simple proportionality of the Nernst effect to the energy dependence $r$ of the scattering time is lost. However, the simple formulae  \eref{nernstdeg} and \eref{nernstnondeg}  should still in most cases give the right order of magnitude, which is what matters for the comparison of compounds in ref.~\cite{Behnia-review-2009}.

%, except in the case of a compensated bands, where the second part vanishes.\cite{Seeger-book-1990}

The degenerate case  \eref{nernstdeg} shows that to find a large Nernst effect one should look for systems with a small Fermi energy (typically not the case in metals) and a large mobility. These are exactly the conditions in bismuth, where there are compensated  electron and hole pockets of small Fermi temperatures, but where the mobility is very large and can exceed $10^{-7}\, \mathrm{cm}^2 \mathrm{V}^{-1}\mathrm{s}^{-1}$. The formula \eref{nernstdeg} applies for an impressive range of parameters, as can been seen in the review by Behnia \cite{Behnia-review-2009} on the Nernst effect in several systems  like NbSe$_2$, heavy fermion systems, 
Bechgaard salts, overdoped cuprates, bismuth to mention some. From results from respective low-temperature regimes ($T\ll T_\fermi$)  Behnia plots $\nu/T$ against $\mu/\erg_\fermi$ \footnote{Note that $r_\mathrm{H}=\mu_\mathrm{H}/\mu \sim 1-2$ for all ordinary scattering mechanisms.\cite{Seeger-book-1990}} for these materials, which turn out to lie impressively close to the line defined by \eref{nernstdeg} with $r$ put to 1 ---over seven orders of magnitude---with bismuth unrivaled at the top.

On the semiconductor side we do not know of any equally complete comparison testing \eref{nernstnondeg} for correlated materials. However, many semiconductors comply with \eref{nernstnondeg} in that  one extracts $r\sim \ordo{1}$.\cite{Sun-private} However, for FeSb$_2$  experiments one extracts  in wide temperature range values of $f$ exceeding 10 and reaching above $10^2$ for certain temperatures.\cite{Sun-private}  Such an energy dependence is difficult to imagine for any ordinary scattering mechanism. So far there is no explanation for this anomalous behavior, but an anomalous density of states and correlations are some possible scenarios.  As mentioned we will not be able to discuss these results with the model we will study. We will instead see how deviations from degenerate analog  \eref{nernstdeg} in principle can come about. Our example uses a  phenomenological model introduced by CST that is  considered to capture electronic correlations.

CST define the symmetry operators $C$ (charge conjugation), $P$ (parity inversion) and $T$ (time reversal) 
\be{threesy,}
\begin{array}{ rcl }
 \chcon^\dagger\elec_{\vec{p}\sigma}\chcon &=& \sigma \elec_{\vect{p}\sigma}^\dagger \\
 \parity^\dagger\elec_{\vec{p}\sigma}\parity &=& \elec_{-\vec{p}\sigma} \\
 \timerev^\dagger\elec_{\vec{p}\sigma}\timerev &=& \elec_{-\vec{p}, -\sigma} \\ 
\end{array}
\ee
where $\parity$ and $\timerev$ are standard, whereas $\chcon$ is not the standard one but a charge conjugation specific to the Fermi surface. For a momentum $\vec{p}=\vec{p}_\fermi+\delta\vec{p}$ above the Fermi surface, the momentum $\vect{p}=\vec{p}_\fermi-\delta\vect{p}$ is pointing in the same direction as $\vec{p}$ but lies just below the Fermi surface such that $\erg_\vect{p}=-\erg_\vec{p}$ where $\erg_\vec{p}=\vec{p}^2/2m-\erg_\fermi$ is the energy with respect to the Fermi surface. 
The hole created by $\elec_{\vect{p},-\sigma}$ has compared to the electron created by $\elec_{\vec{p}\sigma}$ the same spin $\sigma$, the same energy $\erg_\vec{p}>0$, almost the same velocity $\vect{p}/m=\vec{v}_\fermi-\delta \vect{p}$ but almost the opposite momentum $-\vect{p}/m$. 
The free hamiltonian is rewritten as 
\be{}
H_0 \equiv \sum_{\vec{p}\sigma} \erg_\vec{p} \elec_{\vec{p}\sigma}^\dagger  \elec_{\vec{p}\sigma}  
= \sum_{|\vec{p}|>p_\fermi, \sigma }  \Psi_{\vec{p}\sigma}^\dagger\mathbf{H}_{\vec{p}\sigma}\Psi_{\vec{p}\sigma} \, ,
\ee 
where
\be{}
\mathbf{H}_{\vec{p}\sigma}=\mtrx{cc}{\erg_\vec{p} & 0 \\ 0  &  -\erg_\vect{p} } \quad \textrm{and}\quad
\Psi_{\vec{p}\sigma}
\equiv \mtrx{c} {\elec_{\vec{p}\sigma} \\ \sigma\elec_{\vect{p},-\sigma}^\dagger }\, . 
\ee
The upper spinor component accounts for states above the Fermi surface and the lower spinor accounts for states below the Fermi surface. This should be contrasted with the Nambu/Bogoliubov-de Gennes form of the BCS hamiltonian in which the spinor 
$\Psi_\vec{p}=\mtrx{cc} 
{\elec_{\vec{p}\up} & \elec_{-\vec{p}\down}^\dagger }^\mathrm{T}$
has the upper element only creating spin up electron, but unrestricted in momentum, and the lower element creating a particle of positive charge of spin up, but now of the same momentum $\vec{p}$ and oppositve velocity $-\vec{p}/m$. Thus, in the BCS case the two components are related by $\chcon\parity$ conjugation (here with  $\chcon^\dagger\elec_{\vec{p}\sigma}\chcon =  \elec_{\vec{p}\sigma}^\dagger$ ). %As in the BCS case the lower element of $\mathbf{H}$ will be complex conjugated in the presence of a vector potential.
As commented by CST a relaxation mechanism that distinguishes between $\chcon \parity$ components would result in a slow relaxation of longitudinal current but a fast relaxation of the Hall current, which actually seems to occur in superconducting fluctuations above $T_\mathrm{c}$.\cite{Aronov-Hikami-Larkin-1995}

\begin{table}
 \begin{center}
\caption{Transport coefficients derived from the analysis of refs.~\cite{Coleman-Schofield-Tsvelik-PRL-1996,Coleman-Schofield-Tsvelik-1996}, but with generalized temperature dependences $\Gammaf \propto T^\texpf$ and $\Gammas \propto T^\texps$.  The cyclotron frequency $\omegac=eB/m$, the Lorentz number $\mathcal{L}_0=\pi^2 \kbol^2/3e^2 $  and the shorthand notation $\Gamma_\pm=(\Gammaf \pm\Gammas)/2$ are used. 
  Leading temperature behavior given for the case when $\Gammaf \gg\Gammas$ where $\Gamma_\pm \sim \Gammaf/2$.     }
  \begin{tabular}{ccc} 
  \hline
\\
  \begin{tabular}{p{2cm}}
Transport \\ coefficient
\end{tabular}
 &  $\times(e^2n/m)$ & 
 \begin{tabular}{p{2cm}}
Leading  $T$-behavior
\end{tabular} \\
\hline
$\sigma_{xx} $ &  $\Gamma_+^{-1}$  & $T^{-\texpf}$ \\
$\sigma_{xy} $ &  $\frac{\omegac}{\Gammaf\Gammas}$  & $T^{-\texpf-\texps}$  \\
$\jelt_{xx}$  & $\frac{eT\mathcal{L}_0}{2 \erg_\fermi} 
\left( \frac{1}{\Gamma_+}+\frac{\Gamma_+}{\Gammaf\Gammas} \right) $  
& $T^{1-\texps}$ \\
 $\jelt_{xy}$  & $\jelt_{xx}\frac{\omegac}{\Gamma_+}   $  
& $T^{1-\texpf-\texps}$ \\
%$\jhe_{xx}$  & $ T\jelt xx$  &  $ T^{2-\texps}$ \\
%$\jhe_{xy}$  &  $\jhe_{xx}\frac{\omegac}{\Gamma_+} $  &  $T^{2-\texpf-\texps}$ \\
$\jht_{xx}$  &  $T\mathcal{L}_0 \frac{\Gamma_+}{\Gammaf\Gammas} $  &  $ T^{1-\texps}$ \\
$\jht_{xy}$  &  $\kappa_{xx}\frac{\omegac}{\Gamma_+}$  &  $T^{1-\texpf-\texps}$ \\
\hline 
  \end{tabular}
\label{t:cst}
 \end{center}
\end{table}

CST introduce the even and odd eigenmodes under charge conjugation 
\be{}
\begin{array}{rclc}
a_{\vec{p}\sigma } &=& \frac{1}{\sqrt{2}} \left( \elec_{\vec{p}\sigma}+\sigma \elec_{\vect{p},-\sigma}^\dagger \right) \quad\quad & (\chcon=+1)  \\
b_{\vec{p}\sigma } &=& \frac{1}{i\sqrt{2}} \left( \elec_{\vec{p}\sigma}-\sigma \elec_{\vect{p},-\sigma}^\dagger \right)  & (\chcon=-1)\, ,
\end{array}
\ee   
where the index  $\vec{p}$ of these Majorana fermions (here in the general sense that they are real fermions with definite $\chcon$-parity, not that $c^\dagger=c$) is restricted to lie above the Fermi surface to avoid double counting. The Boltzmann equation is transformed into the basis of the Majoranas, in which the distribution function reads 
\be{}
f(\vec{p},\vec{x},t)=
\mtrx{cc}
{\langle a^\dagger a \rangle_{\vec{p},\vec{x},t}  & \langle b^\dagger a\rangle_{\vec{p},\vec{x},t} 
\\
\langle a^\dagger b \rangle_{\vec{p},\vec{x},t}  & \langle b^\dagger b\rangle_{\vec{p},\vec{x},t}     }  
\ee 
In this basis the free evolution, including the EM fields, has  an intricate non-diagonal structure in the Boltzmann equation. However, it is in this basis that CST can easily implement the phenomenological assumption that even and odd components have different relaxation rates by introducing collision integral given by the relaxation time approximation  
\be{}
I[f] = -\frac{1}{2} 
\left\{ 
\mtrx{cc}{\Gammaf & 0 \\ 0 & \Gammas},f-f^\mathrm{eq} 
\right\} 
\ee 
arbitrarily assigning the fast relaxation rate $\Gammaf$ to the $\chcon$-even part and the slow relaxation rate $\Gammas$ to the $\chcon$-odd part. (The opposite choice gives the same results.) In the limit $\Gammaf=\Gammas=\tau_\mathrm{tr}^{-1}$ one recovers the usual relaxation time approximation $I[f]=-(f-f^\mathrm{eq})/\tau_\mathrm{tr}$.  The derivation and solution of the pertinent Boltzmann equation is rather cumbersome and we refer to ref.~\cite{Coleman-Schofield-Tsvelik-1996} for details. From the derived non-equilibirum part  $g=f-f^\mathrm{eq}$ CST  calculate the transport tensors,  given in table~\ref{t:cst}.   In particular, they found that with the choice $\texpf=1$ and $\texps=2$ one could capture the observed temperature dependence of $\sigma_{xx}$ and $\sigma_{xy}$ in cuprate metals. Furthermore, the temperature dependences thereby forced upon   the other transport coefficients are in some cases but not in all in impressive agreement with experiments. 

CST and the subsequent work discuss  also the consequences for the trasport result that can be composed from the coefficients in table~\ref{t:cst}, like the Seebeck coefficient, the magnetoconductivity, the Wiedemann Franz-Law and its Hall version.  However, the main focus was the resulting power law of the temperature dependence and as far as we can see the quantitative enhancement of the signals compared to the standard results was not stressed. CST noted the predicted independence of temperature of the Seebeck coefficient, but were brief on it as they in the cuprate metals  expected a strong modification due to phonon contributions to thermoelectric transport (which could possibly explain the failing agreement of table~\ref{t:cst} with  the thermal coefficients  in the  considered experiments). 
Here we record these derived coefficients, e.g. the Seebeck coefficient
\be{seebeckcst}
S=e\frac{\jelt_{xx}}{\sigma_{xx}}\longrightarrow \frac{eT\mathcal{L}_0}{8 \erg_\fermi} \frac{\Gammaf}{\Gammas} \propto T^{1+\texpf-\texps}\, , 
\ee
that exceeds the Mott result by the ratio $\Gammaf/\Gammas$,
the Wiedemann-Franz ratio, which  is modified into 
\be{}
\frac{\jht_{xx} }{T\sigma_{xx}} = \mathcal{L}_0 \frac{\Gamma_+^2}{\Gammaf\Gammas} \longrightarrow  \frac{ \mathcal{L}_0 \Gammaf }{2\Gammas} \propto T^{\texpf-\texps}\, .
\ee
and is thus enhanced, 
whereas the Hall Lorentz number defined by  $\frac{\jht_{xy} }{T\sigma_{xy}}$ stays interestingly  intact \cite{MeirongLi-PRB-2002}. For the magnetoconductivity $\Delta  \sigma_{xx}=(\sigma_0(B)-\sigma_0)/\sigma_0(B)$ for which one has to go beyond the linear order in $B$ of table~\ref{t:cst} (however, only to quadratic order for the weak field result)  we find the result 
\be{}
\Delta \sigma_{xx}  = -\frac{\omegac^2}{4} \left (\Gammas^{-1}-\Gammaf^{-1} \right )^2 
%\longrightarrow  \sim T^{-2 \texps} 
\ee
which has the same asymptotic temperature dependence as the one presented by  CST but differs most importantly in the relative sign. Only our result vanishes when $\Gammaf=\Gammas$ as expected for an electron gas with an energy independent scattering time.\cite{Seeger-book-1990}

Now we come to the consequences for the Nernst signal, our primary goal.  It appears that this has not at all been discussed in the context of the CST model, neither by CST themselves nor in the subsequent citing literature. However, the result is particularly interesting. With the results in table~\ref{t:cst} inserted into eq.~\eref{nernst} we find 
(assuming $\sigma_{xy}\ll\sigma_{xx}$ ) 
\be{nernstcst}
N =
%\frac{ \omegac S}{e} \left (
%\frac{\Gamma_+}{\Gammaf\Gammas}-\frac{1}{\Gamma_+}
%\right ) \longrightarrow \frac{ \omegac S}{2e} 
%\frac{1}{\Gammas}
\frac{|e| T \mathcal{L}_0}{2 \erg_\fermi} \frac{\omegac}{\Gamma_+} \left (\frac{\Gamma_+^4}{\Gammaf^2\Gammas^2} -1 \right ) 
\longrightarrow  \nonumber \\
\longrightarrow \frac{\omegac |e| T \mathcal{L}_0}{16 \erg_\fermi} \frac{\Gammaf}{\Gammas^2}   
\propto  T^{1+\texpf-2\texps}\, ,       
\ee
to the right in the limit $\Gammaf \gg \Gammas$. In the opposite limit $\Gammaf=\Gammas$ we recover $N=0$, that is, Sondheimer cancellation. 
In the former limit we see that Sondheimer cancellation is avoided even though the relaxation rates were taken to be energy independent. In particular, using that  
\be{}
\mu_\mathrm{H}\equiv \frac{1}{B} \frac{\sigma_{xy}}{\sigma_{xx}}= \frac{\omegac}{B} \frac{\Gamma_+}{\Gammaf\Gammas} \longrightarrow \frac{\omegac}{2 B \Gammas} 
\ee
we find 
\be{nernstmucst}
\frac{\nu}{\mu_\mathrm{H}} \longrightarrow \frac{|e|T\mathcal{L}_0}{\erg_\fermi}\frac{\Gammaf}{8\Gammas}\, . 
\ee
Compared to the result for a degenerate electron gas in eq.~\eref{nernstdeg} we see that the exponent  $r$ of the energy dependence of the relaxation time, typically satisfying $r\sim \ordo{1}$ for ordinary types of scattering mechanisms, has been replaced by the fraction $\Gammaf/\Gammas$ which for $\Gammaf \gg \Gammas$ could substantially exceed $\ordo{1}$.  

Concerning the sign, note that the dominant contribution to \eref{nernstdeg} comes from the term containing the Peltier coefficient $\jelt_{xy}$ whereas in \eref{nernstcst} the term containing $\jelt_{xx}$ dominates, hence the opposite sign in \eref{nernstmucst}. 

In the present model the relaxation rates $\Gammaf$ and $\Gammas$ were assumed to be energy independent. A possible generalization would be to make them energy dependent. This would alter the total temperature dependence of these relaxation rates, which however can be reset as they are anyway chosen phenomenologically. Apart from that there could be a modification of numerical prefactors depending on the exponents $r$, however, in contrast to the ordinary case $\Gammaf=\Gammas$ where prefactors are important and  cancellations can come about for  $r=0$ we do not see that happening for $\Gammaf \gg \Gammas$ in the quantities we have studied. Thus, making the relaxation times energy dependent has little to add.

In conclusion, we have have shown that the phenomenological Boltzmann model by Coleman, Schofield and Tsvelik, which is thought to capture some (yet not determined) electron correlation effects, would predict an electronic contribution to the Nernst effect that could be substantially enhanced compared to the Nernst signal of the Sondheimer formula derived for an degenerate electron gas. Existing results show that correlated metallic systems in the low-temperature regime comply with the ordinary Sondheimer formula for an uncorrelated degenerate electron gas, even when the Nernst signal is huge, and that correlations therefore do not have a direct influence on the Nernst effect. Thus, to date there does not seem to be experimental results that prompt a discussion of the kind of deviations of the Sondheimer formula that we derive. In semiconductors, out of reach of the considered model, the situation is different. The experiments on FeSb$_2$ that  were our initial inspiration display an enhancement of the Nernst effect that cannot be understood with the nondegenerate version of Sondheimers formula given in \eref{nernstnondeg}. Here  correlations could be one of many possible explanations. It would be interesting to see if a semiconductor version of the CST model could be constructed for the non-degenerate case of a semiconductor with   electron and hole bands. Conversely, it would be interesting to see Nernst effect measurements on the doped metallic compound Fe(Sb$_{1-x}$Te$_x$)$_2$ to see if they inherit anything of the anomalous properties of the parent compound.

\acknowledgments
\noindent
 I am grateful to Peijie Sun at MPI-CPfS, Dresden, for many interesting discussion and for showing his unpublished data,  to Kamran Behnia for valuable comments on the manuscript,  to IIP-UFRN, Natal,  for hosting me during a part of the project, and most of all to Stefan Kirchner for introducing me to Peijie and to the problem and for bringing the CST paper to my attention.

\end{document}